\documentclass[prl,aps,amsfonts,amssymb,floats,
twocolumn,showpacs]{revtex4}
\usepackage{epsf}
\usepackage{graphicx}

\begin{document}

\title{Quantum correlations from Brownian diffusion of chaotic level-spacings}

\author{S.N. Evangelou\footnote{e-mail: sevagel@cc.uoi.gr}
and D.E. Katsanos}
%\address
\affiliation{Department of Physics, University of Ioannina,
Ioannina 45110, Greece}

\begin{abstract}
 Quantum chaos is linked to Brownian diffusion
 of the underlying quantum energy level-spacing sequences.
 The level-spacings viewed as functions of their order
 execute random walks which imply uncorrelated
 random increments of the level-spacings
 while the integrability to chaos transition becomes
 a change from Poisson to Gauss statistics for the
 level-spacing increments.
 This universal nature of quantum chaotic spectral
 correlations is numerically demonstrated for eigenvalues
 from random tight binding lattices and for zeros of
 the Riemann zeta function.
\end{abstract}


\maketitle

\par
\medskip
Over the last couple of decades a general consensus has emerged
concerning remnants of classical chaos in quantum physics. Since
the smooth and reversible quantum evolution allows no chaos what
is nowadays called quantum chaos had to be traced elsewhere,
usually in the  stationary eigenvalues of a quantum system
\cite{1}. A key quantity which was introduced to distinguish
between quantum chaos and quantum integrability is the
distribution function of level-spacings between nearest
eigenvalues \cite{1,2}. The quantum chaotic behavior is then
defined as the resemblance of the level-spacing statistics to
those obtained from Gaussian random matrices having highly
correlated levels, which are described by random matrix theory
(RMT) \cite{2,3,4,5}. The major characteristic of quantum chaos is
the so-called ``level-repulsion'' between energy levels which is
due to the dramatic lowering of symmetries in chaotic systems,
while the corresponding level-spacing distribution obeys the
Wigner surmise. On the contrary, quantum systems with integrable
classical analogues are described by Poisson level-spacing
distribution function which implies uncorrelated random
level-spacings with possible ``degeneracies'' related to the
presence of symmetries. Quantum chaos has been experimentally
observed for the irregular energy spectra in atom-optics billiards
\cite{6,7}.

\par
\medskip
The present era of quantum computation, where the exploitation of
quantum information resources came to the forefront, requires a
better understanding of spectral noise arising from quantum
chaotic systems. This noise could be related to quantum
correlations, such as entanglement which is measured via von
Neumann entropy and usually violates Bell inequalities \cite{8,9}.
For such a purpose the energy level-spacing statistical
distribution function is insufficient to describe fully a quantum
system. In order to reveal more about the nature of quantum
spectral noise one must examine level-to-level correlations. This
was done in \cite{10} where they conjecture that quantum chaotic
energy spectra are characterized by $1/f$ noise, which is also
verified by numerical data \cite{10}. In this paper we do not look
only correlations between the enrgy eigenvalues but also view the
nearest level-spacings as functions of their order, so that they
can imitate discrete time series with their order corresponding to
time. We provide evidence that in quantum chaos the classically
chaotic temporal evolution is replaced by diffusion of the
level-spacing dynamics. In other words, we can interpret the
evolution of level-spacings (not the energies themselves) as that
of fictitious particles executing random walks in one-dimensional
space. Moreover, we arrive at a more precise definition of quantum
chaos which  includes quantum correlations in addition to
statistical behavior.

\par
\medskip
The energy level series of a quantum system make up an irregular
continuous but non-differentiable self-affine fractal curve
described by a fractal dimension $D$. For quantum chaotic levels
the noisy curve is space filling with $D=2$ while for energy
levels from quantum integrable systems the absence of correlations
implies less irregular curves with fractal dimension $D=1.5$. The
corresponding power-spectra $P(f)\propto f^{-\alpha}$ are
characterized by a power-law exponent $\alpha$ which obeys the
dimensional relation $D=(5-\alpha)/2$ \cite{11}. Therefore,
quantum chaos implies $1/f$ spectral noise with $\alpha=1$ while
quantum integrability gives $1/f^{2}$ noise with $\alpha=2$,
respectively \cite{10}. In order to provide a better understanding
for the nature of correlations in quantum chaotic energy spectra
we ask the question:  ``What is the nature of correlations  when
dealing with the level-spacings, instead?". Our main finding is
that the correlations in the level-spacing sequences of quantum
chaotic spectra are Brownian. Nevertheless, these are quantum
correlations which occur in addition to the unpredictability
associated to the classical chaotic behavior. Therefore, a link
can be established between quantum chaos and quantum correlations
such as entanglement \cite{8}. These facts are illustrated for two
basic examples of quantum chaotic behavior, the weakly disordered
lattice and the Riemann zeros. Furthermore, quantum chaos is
related to a Gaussian distribution for the $\it
{level}$-$\it{spacing}$ $\it{increments}$ while quantum
integrability is tight up to a Poisson distribution, respectively.
This allows a simple description of the integrability to chaos
transition as a change of the level-spacing increments from the
discrete rare event Poisson distribution to the continuous Gauss
distribution.

\par
\medskip
\begin{figure}
\includegraphics[width=8cm]{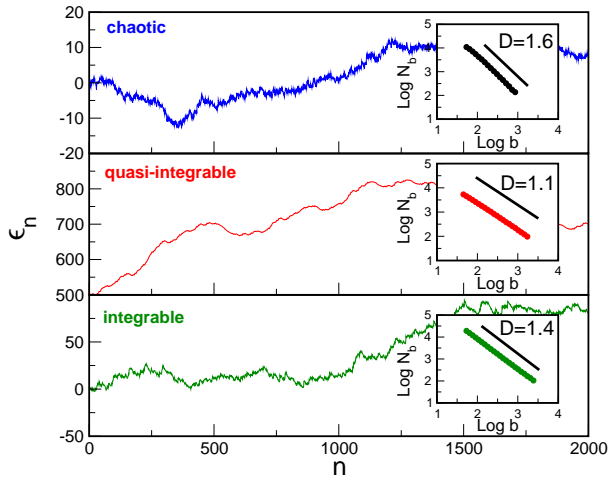}
\caption{(Color online) Energy-level sequences $ {\varepsilon}_{n}$ with
subtracted mean spacings for chaotic, integrable and
quasi-integrable systems obtained from disordered and
quasi-periodic lattices. The decreasing irregularity of the curves
can be seen with the space filling chaotic curve approaching
$D=2$, the integrable curve approaching $D=1.5$ and $D$ close to
$1$ for the chosen quasi-integrable model. The dimensions shown in
the figure differ from the expected values since they were
obtained by box-counting algorithms which are usually not very
accurate for such self-affine curves. Quantum correlations exist
in (a) for the maximally noisy quantum chaotic curves. The quantum
correlations die out in (b),(c).}
\end{figure}

\par
\medskip
In Fig. 1 energy level signals from electrons in lattices  as a
function of the level-index $n$ are displayed. The $n$th level is
expressed as a sum of level-spacings, beginning from the left end
of the spectrum up to the $n$th level, via
\begin{eqnarray}
{\varepsilon}_{n}=\sum_{i=1,n} \left(S_{i}-\langle S \rangle
\right).\nonumber
\end{eqnarray}
From each level-spacing $S_{i}$ we have subtracted a local mean
level-spacing $\langle S\rangle$ by following the usual unfolding
procedure for the density of states to become constant on average
everywhere in the spectrum. The chaotic, quasi-integrable and
integrable energy-level series can be seen to display very
different curve roughness. Very rough curves with $D$ approaching
$2$ are obtained for chaotic levels, the roughness is smaller with
$D$ close to $1.5$ for integrable ones while it becomes much
lower, closer to $D=1$, for the studied quasi-integrable system
where the corresponding curve approaches a line. In Fig. 2 the
corresponding Fourier analysis of these energy-level signals is
shown via the power-spectrum
\begin{eqnarray}
P(f) & = & {\frac {1}{N}} \left|\sum_{n=1}^{N}
{\varepsilon}_{n} \exp\left (-i{\frac {2\pi f n}{N}}\right)
\right|^{2}, \nonumber
\end{eqnarray}
averaged over many configurations. We show the energy levels from
weakly disordered solids and Riemann zeroes which give $1/f$
noise. Our findings for the presence of $1/f$ power-spectra for
quantum chaotic levels agree with the conjecture of Ref.
\cite{10}. We emphasize that quantum correlations exist only for
the maximally noisy spectrum of quantum chaotic levels of Fig.
1(a). The integrable system of Fig. 1(c)  does not show any
correlations described by a less rough curve with Brownian $1/f
^{2}$noise. The quasi-integrable system of Fig. 1(b) does not
display any quantum correlations either. Below we shall
demonstrate that Brownian noise describes quantum correlations for
chaotic systems not for the energy levels but for the
level-spacing sequences, instead.

\begin{figure}
\includegraphics[width=8.0cm] {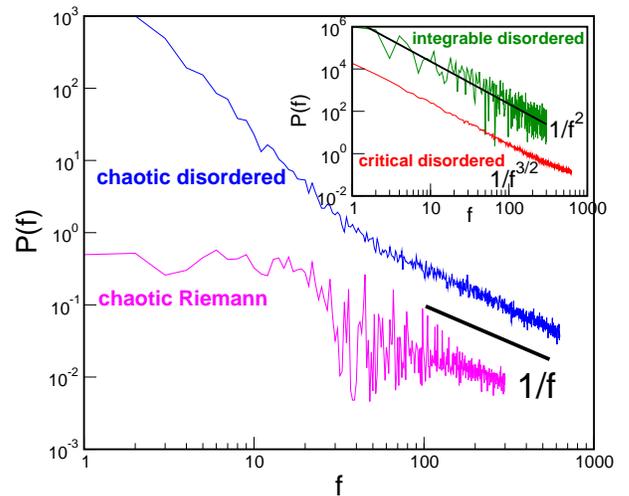}
\caption {(Color online) The averaged power-spectra of the energy levels for
weakly disordered three-dimensional lattice systems, for a
quasi-integrable critical model and for the spectrum of zeros of
the Riemann zeta function. }
\end{figure}

\par
\medskip
The $1/f$ noise found to describe the quantum chaotic levels is a
common law for many classical complex phenomena where very rare
intense effects require the cooperation of various favorable
cases. It can be also viewed as a scaling law since a change of
the intensity combined with contraction in time leaves its form
unchanged. The $1/f$ law is very old and appears in many cases
ranging from fluid flows, stock market price indices, earthquakes,
etc. \cite{12}. However, despite of its generality its origin
remains a mystery without any widely accepted general mechanism
for its interpretation. The most common explanation of $1/f$ noise
relies on the extension of Lorentzian correlation function
$\gamma/(\omega^{2}+\gamma^{2})$ from usual random telegraph noise
of a single fluctuator with exponential relaxation $\gamma$, to
several fluctuators with a broad distribution of relaxation rates.
Although hard to justify physically if $log(\gamma)$ is uniformly
distributed the average over Lorentzians with the chosen broad
distribution of $\gamma$'s gives $1/f$ noise \cite{13}. Another
mechanism of $1/f$ noise seems more appropriate for quantum
chaotic levels. It was proposed (see for example Ref. \cite{14})
that a possible origin of $1/f$ noise can be Brownian correlated
${\it interevents}$ between narrow pulses. Therefore, in order to
obtain $1/f$ noise for the chaotic levels $\varepsilon_{n}$  the
corresponding level-spacings $S_{n}$ which appear in Eq. (1)
should follow a Brownian random walk and the $\it {level}$-$\it
{spacing}$ $\it {increments}$ (spacings of spacings)
$S_{n}-S_{n-1}$ to be random uncorrelated having white type of
noise. This can be understood rather easily since Brownian
$1/f^{2}$ noise of the levels arises from uncorrelated $f^{0}$
white noise of the level-spacings, similarly $1/f$ noise observed
for chaotic levels can arise from uncorrelated white noise
level-spacing increments.

\par
\medskip
\begin{figure}
\includegraphics[width=8cm]{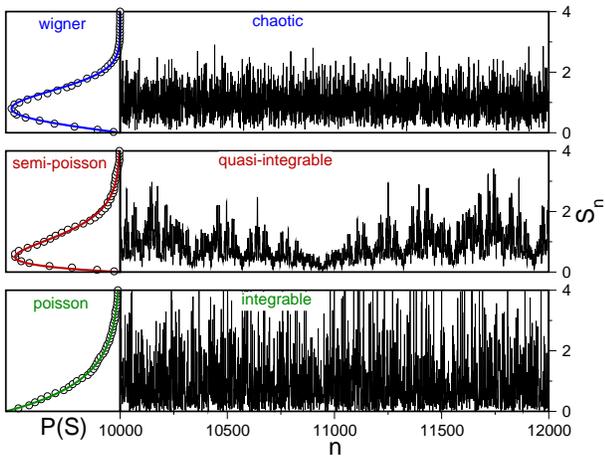}
\caption { (Color online) Level-spacing sequences $S_{n}$ with unit mean-level
spacing $\langle S\rangle=1$  for chaotic, quasi-integrable and
integrable systems. On the left the corresponding normalized
level-spacing distribution function $P(S)$ can be seen.}
\end{figure}

\par
\medskip
Our main results for the level-spacing series
$S_{n}=\varepsilon_{n}-\varepsilon_{n-1}$ are displayed in Fig. 3
together with the corresponding level-spacing distributions $P(S)$
on the left of the figure. It is well-known that from the chaotic
series the obtained $P(S)$ obeys the Wigner surmise, the less
irregular integrable series is described by a Poisson
distribution, while the quasi-integrable ones follow an
intermediate semi-Poisson distribution \cite{15}. The difference
between the amount of noise displayed in the various cases is
immediately obvious. In order to show that quantum correlations
are included in the chaotic level-spacings which follow Brownian
motion we have looked at the statistical distribution of the
level-spacing differences $S_{n}-S_{n-1}$. For chaotic levels it
should be Gaussian and for integrable levels of Poisson type which
correspond to correlated and uncorrelated levels, respectively.
The two behaviors are clearly distinguished in Fig. 4. We have
verified this picture by computing all the levels, the
level-spacings and the increments of level-spacings in lattice
models and the Riemann zeros. In Fig. 4 we show the statistics of
the level-spacing increments $S_{n}-S_{n-1}$. For weakly
disordered lattices where $\beta=1$ and $30000$ Riemann zeroes
\cite{16} where $\beta=2$ the data can be described by simple
Gauss distributions for both cases. In the integrable case the
statistics is Poisson while a similar result of different Poisson
type is obtained for the quasi-integrable statistics in the inset.

\par
\medskip
%\section{Quantum chaos transition}

\par
\medskip
%\subsection{ Disordered tight-binding lattice}

\begin{figure}
\includegraphics[width=8.0cm] {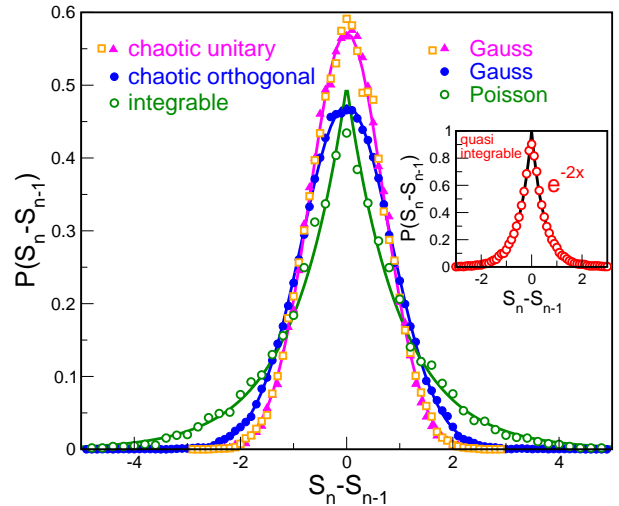}
\caption {(Color online) The distribution of the level-spacing increments.
The transition from integrability to quantum chaos becomes a
transition from Poisson to Gauss. For weak disorder the obtained
variance of the Gaussian is $\sigma^{2}\approx 0.71$ and for the
Riemann zeros is $\sigma^{2}\approx 0.48$. The quasi-integrable
system shown in the inset is described by the simple Poisson
$e^{-2x}$.}
\end{figure}

\par
\medskip
We summarize our numerical computations of eigenvalues responsible
for the level-spacings. First, we consider the weakly disordered
Anderson model studied on $L^{3}$ sites of a 3D lattice. The
diagonal matrix elements of the corresponding are random taken
from a uniform probability distribution of zero mean and variance
$W^{2}/12$ and the off-diagonal matrix elements connecting nearest
neighbors are unity. The short ranged and sparse real symmetric
matrices are diagonalized to obtain long enough signals of raw
eigenvalues $E_{n}$. For such purpose the Lanczos algorithm was
employed and the scaling behavior is achieved by considering
linear lattice size up to $L=25$ for many random configurations.
In the unfolding procedure one replaces each level-spacing
$E_{n+1}-E_{n}$ by
$\varepsilon_{n+1}-\varepsilon_{n}={\cal{N}}(E_{n+1})-{\cal{N}}(E_{n})$,
where $\varepsilon_{n}={\cal{N}}(E_{n})$ are the unfolded levels,
with ${\cal{N}}(E)$ the averaged integrated density of states. In
this procedure the mean level-spacing  becomes $\langle S \rangle
=1$ throughout the spectrum. We have taken linear size $L=25$ for
the chaotic levels disorder $W=10$ for $57$ random configurations
total of $147000$ eigenvalues, for the critical levels disorder
$W_{c}=16.5$ for $80$ random configurations total of $140000$
eigenvalues and the integrable case disorder $W=30$ total of
$10763$ random configurations total of $102800$ eigenvalues. The
averaged power spectra which are shown in Fig. 2 for an energy
window $[0,2]$ which include approximately $2000$ eigenvalues. We
have estimated the power spectrum laws $\alpha=2.0\pm 0.07$ for
the chaotic case, $\alpha=1.52\pm 0.04$  for the integrable case
while a law close to $\alpha=1.62\pm 0.08$ was found for the
critical case (not shown).

\par
\medskip
We have also diagonalized a simpler critical model replacing the
critical case of 3D disordered systems to  consider the behavior
at the borderline from integrability to chaos. This is the
one-dimensional quasi-integrable Harper model \cite{15} where the
site potential is $V_{n}=2\cos(2\pi\sigma n)$ with irrational
$\sigma =(\sqrt{5}-1)/2$. This model displays critical behavior
with multifractal spectra and wave functions. We find that the
averaged power spectra gave power-law estimates close to
$\alpha=3.3$ to $3.8$ as the system size increased.
%\par
%\medskip
%\subsection{Zeroes of the Riemann zeta function}
%\par
%\medskip
The level-spacings of adjacent Riemann complex zeroes \cite{16}
also exhibit fluctuations similar to the ones of the energy
spectra of electrons in disordered lattices. The Riemann
hypothesis states that all non-trivial zeros of the Riemann
$\zeta(s)=\sum_{s}n_{s}$ function have real part equal to $1/2$
with $\zeta_{n}=1/2+\varepsilon_{n}$. Although numerical
computations agree with Riemann hypothesis a full theoretical
proof is not known. However, the RMT is believed to be valid in
this case so that the spectrum of Riemann zeros represents a
hypothetical quantum chaotic system in the unitary universality
class described by GUE, which corresponds to broken time-reversal
invariance via a magnetic field. The $P(S)$ statistics for small
$S$ in this case is $\propto S^{\beta}$ with $\beta=2$, compared
to $\beta=1$ for the  orthogonal case of disordered systems. In
RMT language we have a crossover from GOE to GUE statistics. In
Fig. 2 we demonstrate $1/f$ noise also for sequences of the
Riemann zeroes. We summarize that our findings for fractal
fluctuations and $1/f$ noise for the chaotic spectra are linked to
quantum fluctuations.

\par
\medskip
It is reasonable to expect that fingerprints of chaos present in
the classical time evolution can be also found in the stationary
energy levels of the corresponding quantum system. In addition to
classical chaos  the chaotic spectra should also contain some form
of quantum correlations. Therefore, it is worth searching for the
two ingredients of quantum chaos: ``correlations" for the quantum
part and ``randomness" for the chaotic part. We find that both
appear in the self-affine fractal noise of the spectral curves
which is exemplified via the random walk behavior found for the
level-spacings. Although quantum evolution is very different from
that of classical point particles in phase space Brownian
diffusion also occurs in the quantum case. In Hilbert space
level-spacing evolution is important, which seems to play the same
role for quantum chaos as the time evolution in phase space for
classical chaos. Moreover, quantum chaos implies random
uncorrelated level-spacing increments which obey a Gaussian
distribution. In other words, quantum memory effects in the
evolution, such as dependence on the initial wave packet
superpositions, eventually die out for the level-spacing
increments where quantum chaos meets classical chaos. The quantum
chaotic level-spacing increments obey Gauss distribution
distinguished from integrable ones which obey Poisson
distribution.

\par
\medskip
We have resolved numerically important issues concerning  the
nature of spectral correlations and noise in quantum chaos. Common
signatures of quantum chaos such as ``level-repulsion" and
``spectral rigidity" arise from quantum ``correlations" which
appear in chaotic level-spacings. If combined with ``randomness",
which might have a classical chaotic origin, make up what we call
quantum chaos. On the other hand, integrability usually concerns
uncorrelated random levels where the level-spacings obeys the
Poisson distribution. We should add that integrability occurs
either for pure ballistic systems where quantum correlations exist
without randomness  or strongly disordered systems where due to
localization in random positions randomness exists without
correlations since localized wave functions do not communicate.
Quantum chaos requires both ingredients, ``quantum correlations"
and ``randomness", simultaneously, for example both of them are
present in RMT spectra.

\par
\medskip
In summary, starting from the identification of proposed $1/f$
spectral fluctuations in quantum chaos numerical evidence was
given in favor of random walk behavior for the level-spacings. We
also argued that the definition of quantum chaos requires both
``quantum correlations" and ``randomness". Our numerical study
suggests a connection between chaotic motion of classical
particles in phase space and motion of the level-spacings between
eigenvalues in Hilbert space. In the quantum case Brownian
correlations are shown to contribute to quantum effects. Although
the level-spacing distribution is widely accepted as the best
diagnostic tool of quantum chaos there is no direct theory
attached to it, as it exists for example for other level
correlation functions via RMT (see however the recent study of
\cite{17}). This might explain why the simple random walk behavior
of level-spacings in quantum chaotic systems, where quantum
correlations and unpredictable behavior are combined was not paid
much attention before.

\end{document}